\documentclass[dvipsnames, twocolumn]{aastex63}
\usepackage[version=4]{mhchem}
\usepackage{amsmath}

\received{XX}
\revised{YY}
\accepted{ZZ}
\submitjournal{ApJ}

\shorttitle{The final stage of planet formation chemistry}
\shortauthors{Bosman et al.}

\begin{document}
\title{Destruction of refractory carbon grains drives the final stage of proto-planetary disk chemistry}
\date{\today}

\correspondingauthor{Arthur Bosman}
\email{arbos@umich.edu}

\author[0000-0003-4001-3589]{Arthur D. Bosman}
\affiliation{Department of Astronomy, University of Michigan,
323 West Hall, 1085 S. University Avenue,
Ann Arbor, MI 48109, USA}

\author[0000-0002-2692-7862]{Felipe Alarc\'on}
\affiliation{Department of Astronomy, University of Michigan,
323 West Hall, 1085 S. University Avenue,
Ann Arbor, MI 48109, USA}

\author[0000-0002-0661-7517]{Ke Zhang}
\affiliation{Department of Astronomy, University of Michigan,
323 West Hall, 1085 S. University Avenue,
Ann Arbor, MI 48109, USA}

\author[0000-0003-4179-6394]{Edwin A. Bergin}
\affiliation{Department of Astronomy, University of Michigan,
323 West Hall, 1085 S. University Avenue,
Ann Arbor, MI 48109, USA}

\begin{abstract}
Here we aim to explore the origin of the strong \ce{C2H} lines to reimagine the chemistry of protoplanetary disks. There are a few key aspects that drive our analysis. First, \ce{C2H} is detected in young and old systems, hinting at a long-lived chemistry. Second, as a radical, \ce{C2H} is rapidly destroyed, within $<$1000 yr. 
These two statements hint that the chemistry responsible for \ce{C2H} emission must be predominantly in the gas-phase and must be in equilibrium. Combining new and published chemical models we find that elevating the total volatile (gas and ice) C/O ratio is the only natural way to create a long lived, high \ce{C2H} abundance. 
Most of the \ce{C2H} resides in gas with a $F_\mathrm{UV}/n_\mathrm{gas} \sim 10^{-7}\,G_0\, \mathrm{cm}^3$. To elevate the volatile C/O ratio, additional carbon has to be released into the gas to enable an equilibrium chemistry under oxygen-poor conditions. Photo-ablation of carbon-rich grains seems the most straightforward way to elevate the C/O ratio above 1.5, powering a long-lived equilibrium cycle.  
The regions at which the conditions are optimal for the presence of high C/O ratio and elevated \ce{C2H} abundances in the gas disk set by the $F_\mathrm{UV}/n_\mathrm{gas}$ condition lie just outside the pebble disk as well as possibly in disk gaps. This process can thus also explain the (hints of) structure seen in \ce{C2H} observations.
\end{abstract}


\section{Introduction}
The abundance of the volatile elements (carbon, nitrogen, oxygen and sulfur), and whether these are in the gas, incorporated in ices, or part of the refractory material, is an important parameter in proto-planetary disk physics and chemistry. Volatile elemental abundances influence the molecular composition which in turn influences the ionization and temperature of the gas. Furthermore, giant planets forming in the disk will accrete the local gas, so the elemental composition of the gas will influence the final (elemental) composition of planet. 

The abundance of volatile elements, both in the gas and in the ice, in proto-planetary disks atmospheres appears to be different from the volatile ISM abundances. \textit{Herschel} studies have shown that \ce{H2O} vapor and ice are strongly under abundant, factor 10 to 1000 lower, than expected in ISM composition chemical models in the upper layers beyond snowline \citep{Bergin2010, Hogerheijde2011, Kamp2013, Du2017}. This water is most likely trapped in ice on large grains that have settled to the midplane \citep[e.g.][]{Krijt2016Water}. Furthermore sub-millimeter studies are showing that CO isotopologue emission is weaker than expected \citep[e.g.][]{Favre2013,  Ansdell2016, Miotello2017}. This indicates that the CO abundance is reduced from its expected value ($\sim 10^{-4}$ relative to H$_2$) in the surface layers of the outer ($\gtrsim$ 20 AU) disk. As such the dominant carriers of volatile oxygen and carbon are missing in both the gas and the ice from the surface layers of the outer disk.

As disk mass estimates are usually based on the dust mass, these low oxygen and carbon abundances could be interpreted as a low gas-to-dust ratio. However, Hydrogen-Deuteride (HD) observations towards a handful of disks provide an independent measurement of the gas mass finding gas-to-dust ratios in agreement with earlier assumptions \citep{Bergin2013, McClure2016}. On top of this, measured accretion rates and the composition of accreting material imply disk gas-to-dust ratios that are 100 (ISM) or higher \citep{Kama2015, Manara2016b, McClure2019}. Nitrogen-bearing molecules provide a additional constraint with analysis of \ce{N2H+} and \ce{HCN} also implying gas-to-dust ratios of 100 \citep{vantHoff2017, Cleeves2018, Anderson2019}. Finally, observations of atomic carbon and oxygen lines towards TW Hya are consistent with the missing CO and \ce{H2O} \citep{Kama2016b, Trapman2017}. So it is unlikely that the missing carbon and oxygen are present in unobservable species such as \ce{CH4} and \ce{O2} in upper layers.

At face value, current data and analysis suggest that the carbon and oxygen is sequestered in ices on large grains near the mid-plane. Observational evidence is suggesting that this process is relatively fast and takes place in the first Myr after disk formation \citep[][]{Zhang2020,Bergner2020}.

The low total abundance of CO and the even lower abundance of \ce{H2O} indicate that total volatile C/O ratios are elevated above the ISM ratio of 0.4. This is confirmed by the brightness of the \ce{C2H} lines detected towards many disks \citep{Guilloteau2016, Bergner2019, Miotello2019}, with the \ce{C2H} lines fluxes comparable to \ce{^{13}CO} \citep{Kastner2014}. Models that match these observations require C/O ratios of 1.5 -- 2 \citep{Bergin2016, Miotello2019}. Under these high C/O ratio conditions, CO is the dominant oxygen bearing molecule, so these conditions also explain the low \ce{H2O} fluxes observed with \textit{Herschel} \citep{Kamp2013}. To get to these higher C/O ratios it is not enough to remove volatiles from the surface layers and leave a small fraction of the CO. It is necessary to create a surplus of carbon relative to oxygen. This can be done by extracting oxygen from \ce{CO} and putting it into water ice sequestered near the mid-plane, or by releasing excess carbon from a refractory reservoir, carbonaceous grains or PAHs \citep{Draine1979,Finocchi1997,Visser2007, Alata2014, Alata2015, Anderson2017}. Observations of some galactic PDRs also show the need for elevated C/O ratios \citep[e.g.][]{Guzman2015, LeGal2019}. This high C/O ratio can be caused by release of carbon from grains due to photo-ablation \citep{Alata2015}, a similar process might thus be active in proto-planetary disk surface layers.

Strong \ce{C2H} emission is seen in a majority of proto-planetary disks spread over all ages \citep{Bergner2019, Miotello2019}. Furthermore, \ce{C2H} emission in many disks shows (signs of) structure \citep{Bergin2016, Bergner2019}. As such the conditions for bright \ce{C2H} emission must be set early, persist for the disk lifetime and have some dependence on local disk conditions. The \ce{C2H} chemical time-scales are short, as such {\em the abundant \ce{C2H} thus has to be the result of a long-lived (millions of years) equilibrium cycle.} The goal of this paper is to elucidate the conditions necessary for this cycle.

\section{Chemistry of \ce{C2H}}

\begin{figure*}
    \centering
    \includegraphics[width =\hsize]{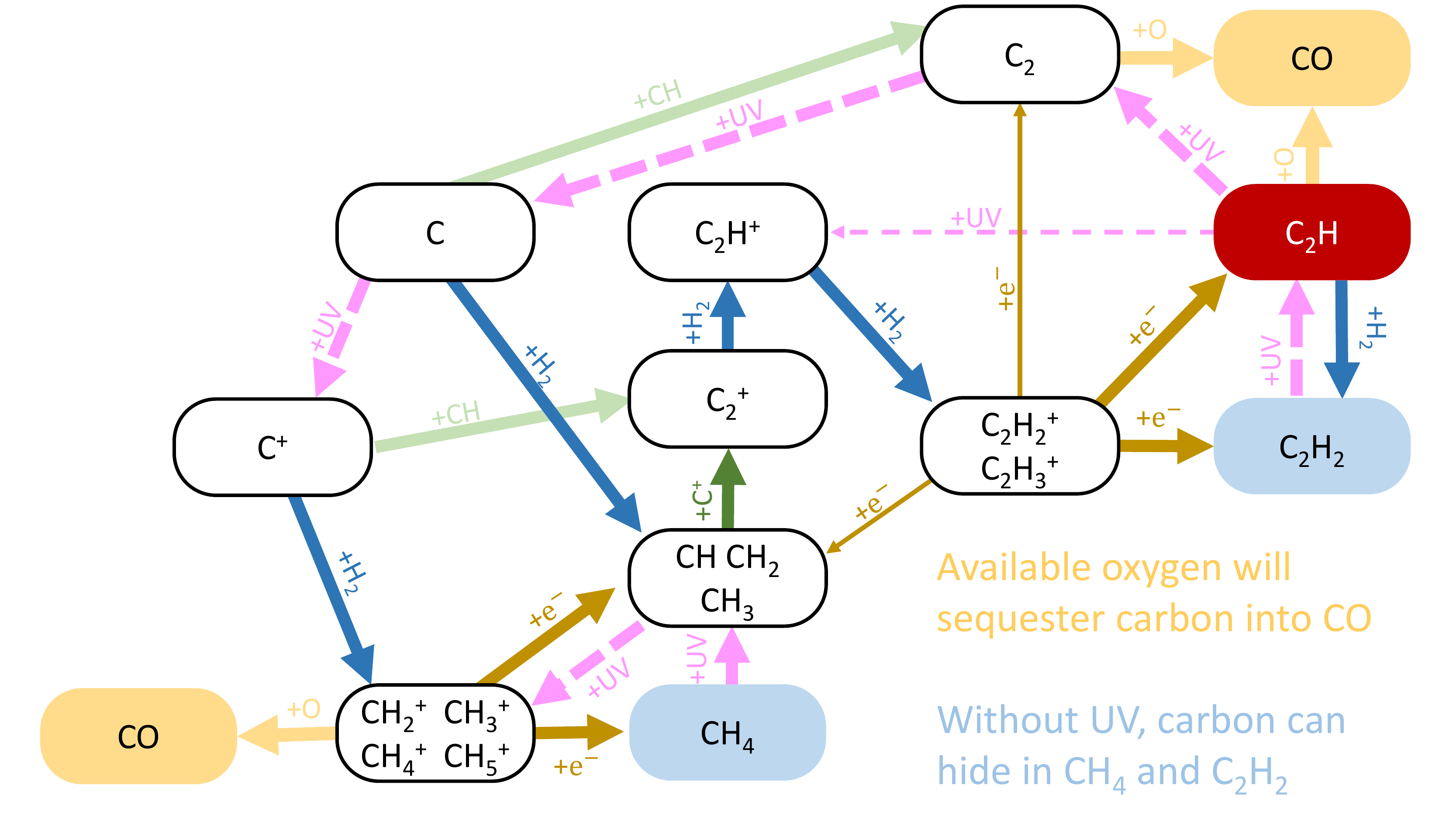}
    \vspace*{0.2cm}
    \caption{\label{fig:reactionnetwork}
    Carbon chemistry network showing the important reaction pathways in the UV dominated layers of proto-planetary disks. Thick arrows show major pathways, thin arrows show minor pathways. Reactions involving oxygen, which always end in CO, are shown in yellow arrows. \ce{CH4} and \ce{C2H2} (light blue background) are species that can only be efficiently destroyed by UV photons. In the absence of UV photons these species can contain a significant amount of carbon. \ce{C2H} (red background) can only be formed if the cycle is active, that is when there are sufficient UV photons to release carbon from \ce{CH4} and \ce{C2H2}. If water ice is present, these same photons would release atomic oxygen from any present \ce{H2O} ice, quenching the \ce{C2H} production, in regions with large amounts of water ice. } 
\end{figure*}

\begin{figure*}
    \centering
    \includegraphics[width =\hsize]{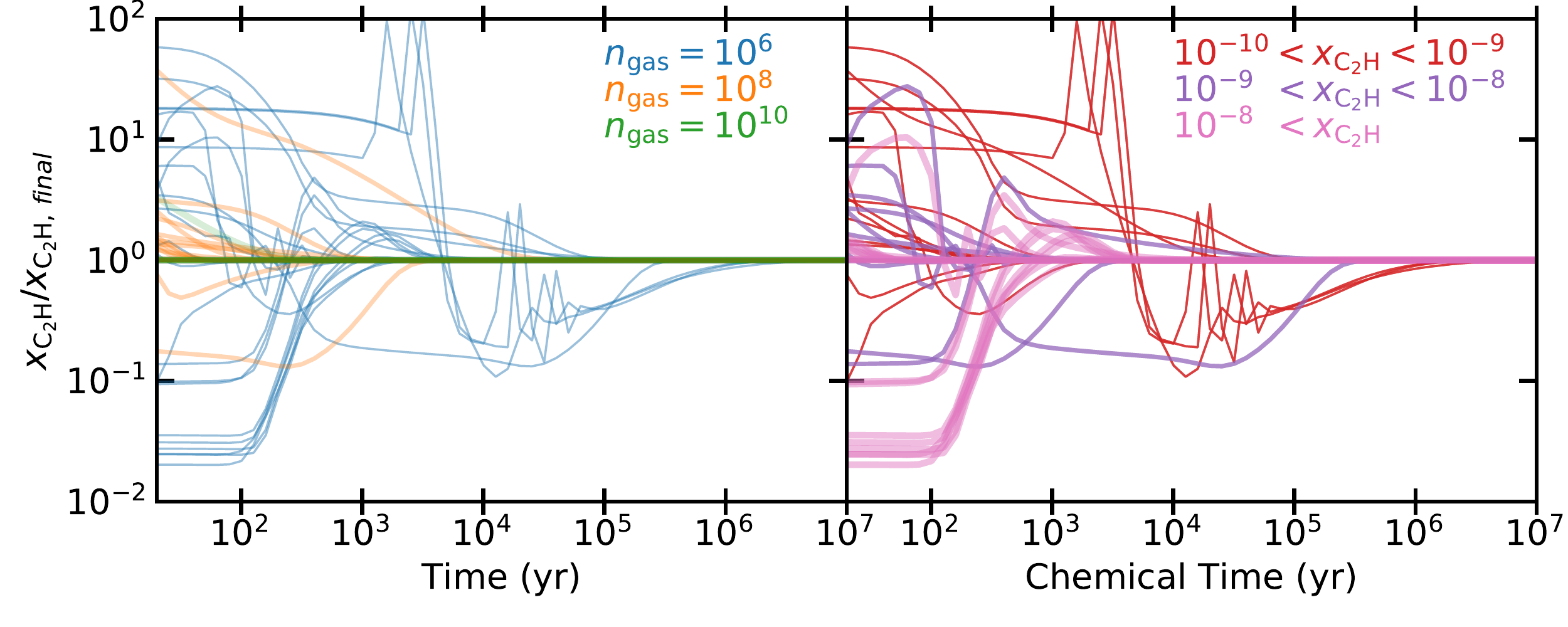}
    \caption{\label{fig:C2Htimescales}
    The \ce{C2H} abundance normalized to the abundance at 10 Myr as function of time using the \citet{Bosman2018CO} gas-grain network. All models that end with a \ce{C2H} abundance greater than $10^{-10}$ are plotted with colors denoting the model densities (\textit{left}) and final \ce{C2H} abundance (\textit{right}). All models converge on the final \ce{C2H} abundance within 1 Myr, and models that have a high \ce{C2H} abundance at the end of the chemical model converge faster than models with a lower abundance. }
\end{figure*}

The Ethynyl radical, \ce{C2H}, is a radical that is often used to trace the C/O ratio of gas \cite[e.g.][]{Bergin2016,Cleeves2018, Miotello2019}. A simplified reaction network for the chemistry that leads to \ce{C2H} is shown in Fig.~\ref{fig:reactionnetwork}. The \ce{C2H} abundance is strongly dependent on the amount of free carbon; that is carbon not contained in CO. Furthermore \ce{C2H} and related hydrocarbons react very quickly, especially with atomic oxygen and the OH radical. Reactions with oxygen bearing species inevitably lead to \ce{CO} as one of the products. As such, to produce a high abundance of \ce{C2H}, a high C/O ratio is necessary \citep{Bergin2016, Miotello2019}. 

As Fig.~\ref{fig:reactionnetwork} shows, to form \ce{C2H} a significant level of UV is also necessary, as further exemplified as its use as PDR tracer \citep[e.g.][]{Jansen1995, Nagy2015}. The level of UV necessary to create abundant \ce{C2H} depends on the density of the gas and the C/O ratio. For a C/O of 0.4 and the low density ($10^{5}$ cm$^{-3}$) outer regions of the disk a $F_\mathrm{UV}  = 1\,G_0$ is enough, while in denser disk surface layers ($10^{9}$ cm$^{-3}$) a $F_\mathrm{UV}  = 10^4\,G_0$ is necessary, where $G_0$ is the \citet{Habing1968} flux of $1.6 \times 10^3$ erg s$^{-1}$ cm$^{-2}$. \footnote{The relation between abundant \ce{C2H} and the physical conditions be explored in more detail at the end of this section.} This means that \ce{C2H} is most abundant in regions with an active chemistry which equilibriates quickly. Figure~\ref{fig:C2Htimescales} shows the converging behavior of the \ce{C2H} abundance in a set of chemical points models for conditions relevant to the disk layers with abundant \ce{C2H}. As UV photons are abundant in the regions where \ce{C2H} is produced, it is not just enough to change the C/O ratio in the gas-phase, for example by freeze-ing out \ce{H2O}. Only by also removing the oxygen from UV dominated layer, so neither photo-desorption nor photo-dissociation can replenish oxygen to the gas-phase, is it possible to increase the \ce{C2H} abundance. 

The chemical models in Fig.~\ref{fig:C2Htimescales} is based on the network presented in \citet{Bosman2018CO} with photo-dissociation reactions from \citep{Heays2017}. The conditions were chosen to encompass the PDR layer of a proto-planetary disk model around a T-Tauri star. The density was varied between $10^6$ and $10^{10}$ cm$^{-3}$ with a UV field between 10 and $10^5$ $G_0$. Gas and dust temperatures were varied between 30 and 300 K. Finally the C/O ratio was varied between 0.4 and 2.0. All models that end up with a \ce{C2H} abundance larger than $10^{-10}$ equilibrate within 1 Myr.

Variations of the initial composition in these chemical models shows, as also shown in \citet{Bergin2016}, that a high initial abundance of \ce{C2H2} or \ce{CH4} will initially lead to a high, $> 10^{-10}$, \ce{C2H} abundances.  However, the high \ce{C2H} abundance is not long lived unless the conditions are right to have a high \ce{C2H} abundance in kinetic equilibrium, which is reached in $10^5$ -- $10^6$ years \citep[Fig. \ref{fig:C2Htimescales} and ][]{Bergin2016}. The physical parameter space in which this happens is bigger when the total C/O ratio is larger than 1. Put simply, there must be more carbon available for the chemistry than oxygen.

\begin{figure}
    \centering
    \includegraphics[width = \hsize]{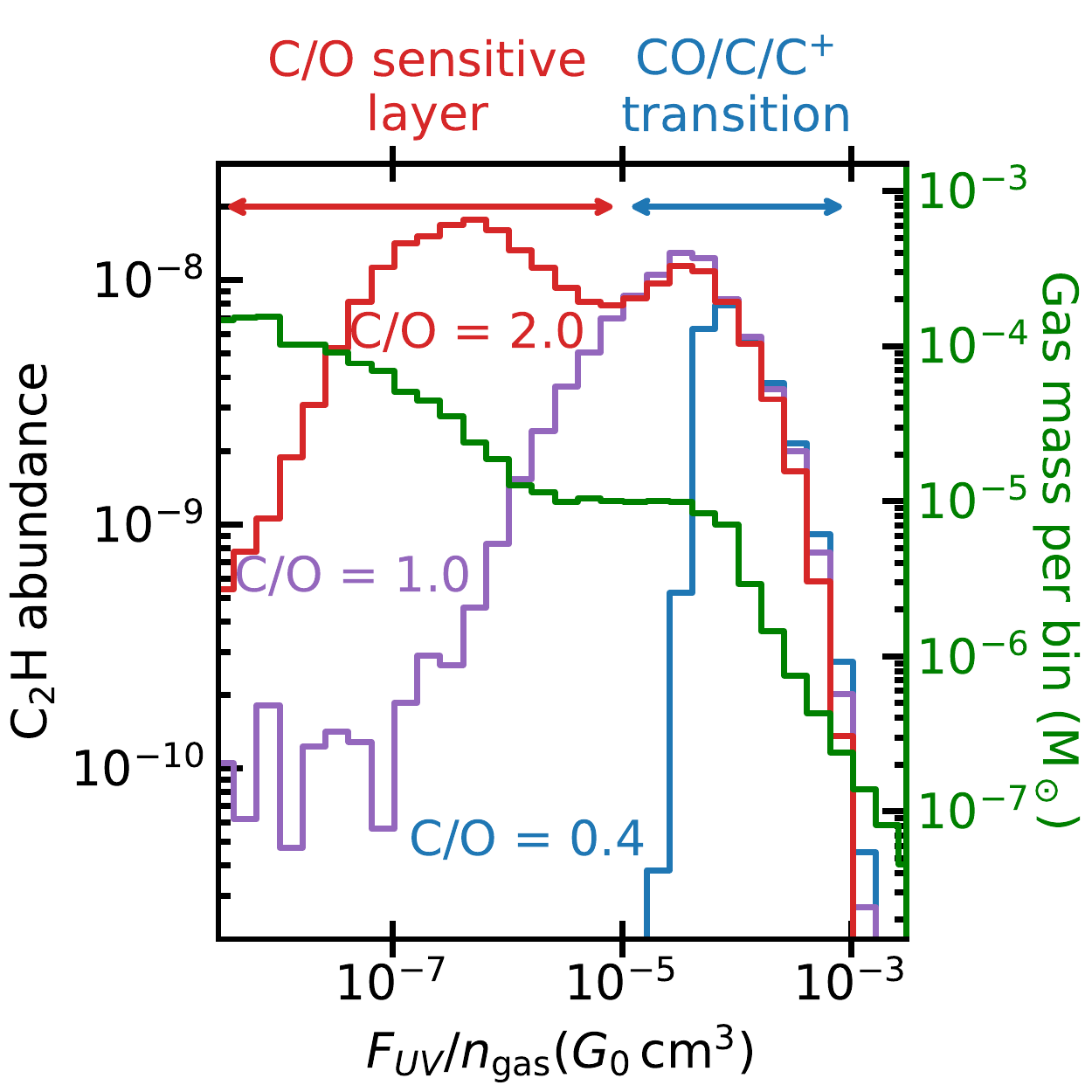}
    \caption{Mass average \ce{C2H} abundance as function of the UV field over the density ($F_\mathrm{UV} / n_\mathrm{gas}$) for a C/O of 0.4 (blue), 1.0 (purple) and 2.0 (red) in the TW Hya model of \citet{Bergin2016} (left axis). The total mass in each of the bins is shown in green (right axis). In regions with $F_\mathrm{UV} / n_\mathrm{gas} > 10^{-5} G_0\,\mathrm{cm}^{3}$ the \ce{C2H} is independent of the \ce{C/O} ratio, only in regions with lower $F_\mathrm{UV} / n_\mathrm{gas}$ does the \ce{C2H} abundance start to strongly depend on the C/O ratio. A smooth disk will naturally have a lot more mass that has lower $F_\mathrm{UV} / n_\mathrm{gas}$, as the density is higher in these regions. This leads to a strong \ce{C2H} abundance increase at higher C/O ratios. }
    \label{fig:C2H_G0_n}
\end{figure}

This can be clearly seen in the \ce{C2H} abundances in the thermo-chemical models from \citet{Bergin2016}. Fig.~\ref{fig:C2H_G0_n} shows the mass averaged \ce{C2H} abundance as function of $F_{\mathrm{UV}}/n_\mathrm{gas}$ for a TW Hya disk model \citep[][]{Bergin2016}. The chemical network used in these thermo-chemical models is different from the one used in the single point models in Fig.~\ref{fig:C2Htimescales}. It contains a more resticted grain surface chemistry and the gas-phase chemistry in RAC2D is based on \textrm{UMIST06} \citep{Woodall2007, Du2014}, while the \citet{Bosman2018CO} model is based on \textrm{UMIST12} \citep{McElroy2013}. The reactions important for \ce{C2H} formation and destruction (see Fig.~\ref{fig:reactionnetwork}) are the same between the two networks.  

$F_{\mathrm{UV}}/n_\mathrm{gas}$ is a central parameter in describing effects on the chemistry and thermal physics within PDR models \citep{Kaufman1999} as it balances destruction and heating ($\propto F_{\mathrm{UV}}$) and formation and cooling ($\propto n_\mathrm{gas}$) rates of molecules, respectively. Through out the paper $F_{\mathrm{UV}}/n_\mathrm{gas}$ will be expressed in units of $G_0\, \mathrm{cm}^3$.

High \ce{C2H} abundances are found between $F_{\mathrm{UV}}/n_\mathrm{gas} = 10^{-8}-10^{-3}\, G_0\,\mathrm{cm}^{3}$, with conditions between $10^{-8}-10^{-5}\,G_0\,\mathrm{cm}^{3}$ are most sensitive to changes in the C/O ratio. As densities in the disk are greater than $ 10^{6}$ cm$^{-3}$, these conditions always correspond to an effective radiation fields $> 0.1\,G_0$.
This effectively means that the gas responsible for \ce{C2H} emission is not fully shielded from UV emission. This supports the conclusion of \citet{Bergin2016} that dust evolution is important for \ce{C2H} formation: dust evolution, especially grain growth and settling increase the penetration of UV in the disk, increasing the mass fraction of the disk where \ce{C2H} can be abundant. Fig.~\ref{fig:C2H_G0_n} also demonstrates that high C/O ratios are needed to elevate the C$_2$H abundance within regions that carry significant mass. As in all our models the maximum \ce{C2H} abundance seems to be around $10^{-8}$, or $\sim$ 0.01\% of the total carbon. In the CO/C/C$^{+}$ transition layer, which is the layer that produces \ce{C2H} at low C/O ratios, this only allows a \ce{C2H} column of order $10^{12} \mathrm{cm}^{2}$. Thus an increase in abundance in the deeper, denser layers of the disk is necessary to reproduce the high \ce{C2H} columns, $10^{14}$--$10^{15}$ cm$^{-2}$ observed, which only happens when the C/O ratio is above 1.0 \citep[e.g.][]{Bergin2016, Bergner2019}. 

\section{Towards a high C/O ratio}
\subsection{High C/O due to Volatile Depletion? }
It is currently unclear what exactly is causing the loss of oxygen and carbon bearing species from the surface and outer regions of protoplanetary disks with leading theories exploring dust evolution or chemical processing \citep{Krijt2016Water, Schwarz2018, Krijt2018, Bosman2018}. The high sublimation temperature of water \citep[150-300~K, depending on density][]{Bergin2018} lends it to be found as water ice for the majority of the disk mass and hence dust grain growth would appear to be most relevant \citep{Krijt2016}.  For CO, the much lower sublimation temperature \citep[20--25 K][]{Schwarz2016, Pinte2018, Qi2019} makes it more difficult for dust evolution to be the sole process and models have therefore also explored chemical processing of CO into less volatile forms such as CO$_2$ or CH$_3$OH \citep{Furuya2014, Schwarz2018, Bosman2018, Dodson-Robinson2018, Schwarz2019}. 

It seems however, that the CO removal process is relatively fast and happens within the first Myr after disk formation \citep{Zhang2020}. This is shorter than the timescales necessary to reduce the \ce{CO} abundance in chemical models using cosmic-ray driven gas-grain chemistry, which are generally $> 1$ Myr, unless elevated cosmic ray ionization rates are assumed \citep[e.g.][]{Schwarz2018, Bosman2018CO}. Furthermore, as \ce{C2H} is created in a photon-dominated layer layer, oxygen carrying species in the ice which are formed from CO destruction, such as \ce{H2O} and \ce{CO2} would be photo-desorbed and dissociated by the same UV that is necessary to form \ce{C2H}. The released oxygen would destroy the \ce{C2H} in the gas-phase. Chemical conversion of \ce{CO} into thus does not directly create the high C/O ratio conditions necessary for the high observed \ce{C2H} abundances. 

A combined chemical-dynamical process would thus be necessary, and the interaction of chemical and dynamical effects seem to strengthen each other, shortening the CO depletion timescale \citep{Krijt2020}. In these models the total C/O ratio is tracked and a rise in C/O ratio is seen. The total, gas+ice C/O does rises to 1.0 between the CO and \ce{CH4} snow surfaces, with a low column layer. In these models the gas-phase C/O ratio is greater than one, but this excess carbon is balanced by \ce{CO2} and \ce{H2O} in the ice. Under the UV conditions necessary for \ce{C2H} production, this oxygen would be released from this ice, and smothering \ce{C2H} formation. These chemical-dynamical models, thus do not create the conditions for strong \ce{C2H} emission naturally.  

Furthermore, if the process of CO depletion is linked with the increase of the C/O ratio above 1.0, there should be a clear trend between CO abundance and \ce{C2H} flux. This, however, is not seen observationally \citep{Miotello2019}. Finally, the \ce{C2H} emission is structured in both TW Hya and DM Tau.  Thus it is likely that the C/O ratio is similarly structured. This is not easily explained by volatile depletion, which seems to be relatively smooth \citep{Zhang2019, Krijt2020}. The depletion of CO is not \textit{directly} responsible for the high C/O ratios necessary to explain the \ce{C2H} abundances. However, the lower elemental abundance of oxygen in the surface layers due to CO depletion does make it easier for a source of additional carbon to elevate the C/O ratio. We therefore propose that the depletion of CO is a necessary pre-condition for the high C/O ratios observed.

\subsection{Photo-ablation of Refractory Carbon}

\begin{figure*}
    \centering
    \includegraphics[width = \hsize]{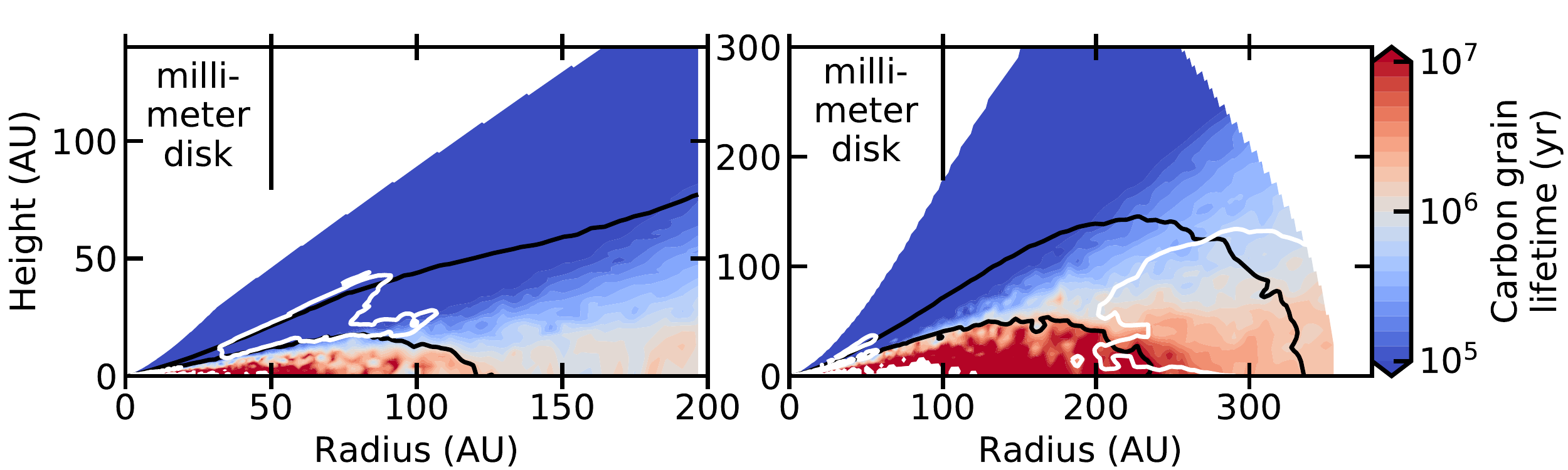}
    \caption{Carbon grain lifetime for the TW Hya (left) and DM Tau (right). Density structure and resulting UV field used for the calculation of the grain lifetime are from the models in \citet{Bergin2016}. Most of the disk surface has a carbon grain lifetime less than 1 Myr, while the region with a refractory carbon lifetime less than 3 Myr spans the entire disk except for pebble disk. The radial size of the pebble disk is denoted with the vertical black line at the top of the figure. The white lines denote the location that \ce{C2H} is abundant, $x_{\ce{C2H}} > 10^{-9}$. The black contours encompass the regions with $F_\mathrm{UV}/n_\mathrm{gas}$ between $10^{-8}$ and $10^{-5}$ $G_0\,\mathrm{cm}^3$. In this region, an increase in C/O ratio leads to the strongest response in total \ce{C2H} abundance. The areas where \ce{C2H} is abundant and where $F_\mathrm{UV}/n_\mathrm{gas}$ is between $10^{-8}$ and $10^{-5}$ $G_0\,\mathrm{cm}^3$ do not overlap completely as the models have a varying C/O ratio. A C/O ratio $<$ 1.0 suppresses \ce{C2H} formation inside 30 and outside 100 AU in TW Hya and between 40 and 200 AU in DM Tau. }  
    \label{fig:carbon_destruction_ts}
\end{figure*}

The initial evolution of the disk leaves the surface layer and outer disk depleted of volatiles and dust. The volatile Carbon and oxygen budget are dominated by \ce{CO} at an abundance between $10^{-6}$ and $10^{-5}$ with respect to \ce{H}. The gas thus has a C/H that is 1 to 2 orders lower than the volatile ISM and a C/O close to unity as a result of the CO and \ce{H2O} depletion episode. Finally the grain growth and settling also allows the UV to penetrate more deeply into the disk. 

If the excess carbon is not drawn from a volatile source, it is thus has to  originate from a refractory source. In interstellar space about 50\% of the carbon is contained in refractory form, from PAHs and nano-particles to amorphous carbon grains and carbon ``goo'' coatings on silicate grains \citep[][]{Greenberg1995, Jones2013Cdust, Chiar2013, Mishra2015}. Carbon can be extracted from these refractory forms by interactions with energetic particles or by oxidation \citep[][]{Draine1979,Finocchi1997, Alata2014, Alata2015, Anderson2017}. As the region of interest here are mostly cold ($<100$ K), and oxygen poor, oxidation should not play a role. As such we will only consider the release of carbon by energetic photons, which can penetrate more deeply into the disk due to the dust evolution. Specifically, we consider the release of carbon from carbonaceous grains due to UV photons, other carbon release mechanisms and carbon reservoirs will be considered in the discussion (Sec.~\ref{ssc:carbon_carrier}). 

Hydrogenated amorphous carbon on the surface of grains can be photo-ablated, releasing the carbon, mostly in the form of \ce{CH4} to the gas-phase \citep{Alata2014}. The grain lifetime, following \citet{Anderson2017}, is given by:  
\begin{equation}
\label{eq:Cgraindestr}
    \tau_{\mathrm{C}_{\mathrm{ref}}} = N_\mathrm{C\,grain}/\left(\sigma Y_\mathrm{C} F_\mathrm{UV} \right)
\end{equation}
where 
\begin{equation}
    N_\mathrm{C\,grain} = \rho_{\mathrm{C}_{\mathrm{ref}}} \frac{4}{3} \pi a^3/m_{\mathrm{C}}
\end{equation}
is the number of carbon atoms per grain, $\sigma$ is the geometric cross section of the grain, $Y_\mathrm{C} = 8 \times 10^{-4}\,\mathrm{photon}^{-1}$ is the carbon sputtering yield \citep[][]{Alata2014, Alata2015}, $F_\mathrm{UV}$ is the UV field, $10^8 \times G_0\,\mathrm{photons}\, \mathrm{cm}^{-2}\, \mathrm{s}^{-1}$, $\rho_{\mathrm{C}_{\mathrm{ref}}} = 2.24\, \mathrm{g}\, \mathrm{cm}^{-3}$ is the density of carbonacous grains, $a = 0.1\,\mu\mathrm{m}$ is the grain radius and $m_{\mathrm{C}}$ is the mass of a carbon atom.
This carbonaceous grain lifetime is calculated for TW Hya and DM Tau disks using the model structures from \citet{Bergin2016} and shown in Fig.~\ref{fig:carbon_destruction_ts}.

The carbon grain lifetime is significantly less than a few million years for most of the \ce{C2H} emitting region. This is less than the expected disk lifetime or even the age of the 5-10 Myr TW Hya disk \citep{Weinberger2013}. Carbonaceous grains can thus be processed enough to remove a significant fraction of the carbon from the grains, enriching the gas. As the carbon grain lifetimes are short, a single enrichment event is expected, in contrast to a slow continuous release of carbon over the disks lifetime. 

It is thus qualitatively possible to enrich the gas with carbon from refractory origin, but what is necessary to quantitatively match the extreme C/O ratios (C/O $\approx 2$)? In the ISM grains contain $\sim $50\% of the total carbon, $10^{-4}$ w.r.t. \ce{H} \citep{Draine2003, Mishra2015}. However, the grains have grown and settled lowering the abundance of small grains, those grains that are well coupled to the gas, increasing the gas-to-dust ratio above 100 in the \ce{C2H} emitting layers. The available carbon abundance in refractory form is thus $10^{-4} \times 100/$gas--to--dust ratio. To elevate the C/O ratio from the depleted state with a C/O of 1.0 to 2.0 it is thus necessary to add as much carbon as there already is in the gas-phase, which is between $10^{-6}$ - 10$^{-5}$ w.r.t. H. This requires ablation of 1-to-10\% of all the refractory carbon originally in the disk surface layers. To have enough grains to provide the carbon it is thus necessary that the gas-to-dust-ratio is $  \leq 1/\left(100 \times \left(C/H\right)_\mathrm{gas}\right)$ in the layers where carbon is efficiently photo-ablated. For strongly volatile depleted disks the surface layer gas-to-dust ratio should thus be less than $10^4$ while for less depleted disks a lower gas-to-dust ratio (around 1000) is necessary to be able to provide the required carbon to the gas.

The actual grain abundance in the surface layers and outer regions of the disk is hard to constrain. SED fitting models generally use a gas-to-dust ratio of 1000-10000 in the surface layers \citep[e.g.][]{Andrews2013}. The TW Hya scattered light model of \citet{vanBoekel2017} also has a gas-to-dust ratio of 10000 in the surface layers, consistent with the SED models, this is a factor of five higher than the gas-to-dust ratio in the thermo-chemical model of \citet{Bergin2016}, which also matches a host of other gas tracers \citep{Du2015}. The DM Tau model of \citet{Bergin2016} even has a gas-to-dust ratio of 125 in the surface layers enough to elevate the C/O ratio to approximately 2.0, even if there is very little carbon and oxygen depletion. These models, while highly degenerate are at least in the right ball park to provide enough grain material in these regions to elevate the C/O ratio to the required values. Settling and drift models show a larger range of dust depletions in the outer disk. Models with a low  $\alpha$ ($< 10^{-3}$), as inferred from observations \citep{Pinte2016,Flaherty2017, Teague2018}, predicting more than a factor 10 depletion of the small dust grains in the surface layers, in general predicting less dust than necessary in SED models \citep[e.g.][]{Facchini2017, Krijt2018, Woitke2019}. 
All these things considered, it is very hard to say what is, and what is not enough turbulence to provide the small grains, and thus excess carbon necessary for the elevated C/O ratios. Disk that have an close to ISM O/H ratio in the surface layers, needs a lot of small dust, consistent with no settling, and thus high very high turbulence ($\alpha > 10^{-2}$). Very oxygen depleted disks, such as TW Hya, however, 1\% of the original dust, which for that disk is consistent with an $\alpha$ of $10^{-4}$ \citep{vanBoekel2017}. 

We note however, only needs a single injection of excess carbon from the small grains. So even if the current levels of small grains, or by extension, current turbulent $\alpha$ is not enough, it is possible that the high C/O ratios are a result of a previous stage of the disk evolution in which there were enough carbonaceous grains in the disk atmosphere.

\section{Discussion}
\subsection{Structure in \ce{C2H}}
Observations of \ce{C2H} show that the \ce{C2H} is structured. High resolution of DM Tau and TW Hya shows an emission ring outside of the pebble disk while lower resolution data from \citet{Bergner2019} and \citet{Miotello2019} hints at structure below the observed resolution in many of the disks observed to date. 

A typical disk models with a tapered power-law surface density structure and a constant gas-to-dust ratio will lead to very smooth \ce{C2H} surface density profiles outside of $\sim$ 20 AU \citep[e.g.][Fig. 7]{Bergin2016}. The structures that are visible are thus due to a combination of radial changes in the C/O ratio and changes in the UV penetration. As the UV penetration can only significantly change the \ce{C2H} abundance for elevated C/O ratios, changes in the C/O ratio are the expected dominant driver in the \ce{C2H} structure. If the C/O ratio is elevated due to the photo-ablation of carbonaceous grains, then the C/O ratio is linked to the (historical) UV penetration and small grain abundances. 

\subsubsection{Rings outside of the pebble disk}
Outside of the pebble disk, the radiation field is a combination of the interstellar radiation field, that can reach a large volume of the outer disk, and stellar UV photons that scatter from the disk surface downward. The radiation field is typically between 0.1 and $10 \times G_0$. As the external UV is only barely attenuated in these disk regions, changing the amount of small dust does little to change the UV field in the outer disk. As such the increase in \ce{C2H} column must be due to a change in the C/O ratio outside of the pebble disk. 

The region outside the pebble disk is a natural place for elevated C/O ratios to occur due to photo-ablation. Grain coagulation is less effective at larger disk radii as densities are lower \citep[e.g.][]{Brauer2008}.  With less growth, a smaller fractions of grains will have settled, leaving a larger reservoir of small grains ($a < 0.1 \mu$m) exposed to UV photons. On top of that, vertical mixing is expected to concentrate volatiles in ices on pebbles near the disk mid-plane, as there is a lack of pebbles near the mid-plane this volatile sink is not present, and the excess carbon can potentially stay in the gas-phase for the entire disk lifetime. Of course, this does depend on timescales for radial motions and gas loss via winds.  However, beyond the pebble disk it is clear that the depletion cycle will not be activated.

\subsubsection{Structures in the Pebble Disk}
\begin{figure}
    \centering
    \includegraphics[width = \hsize]{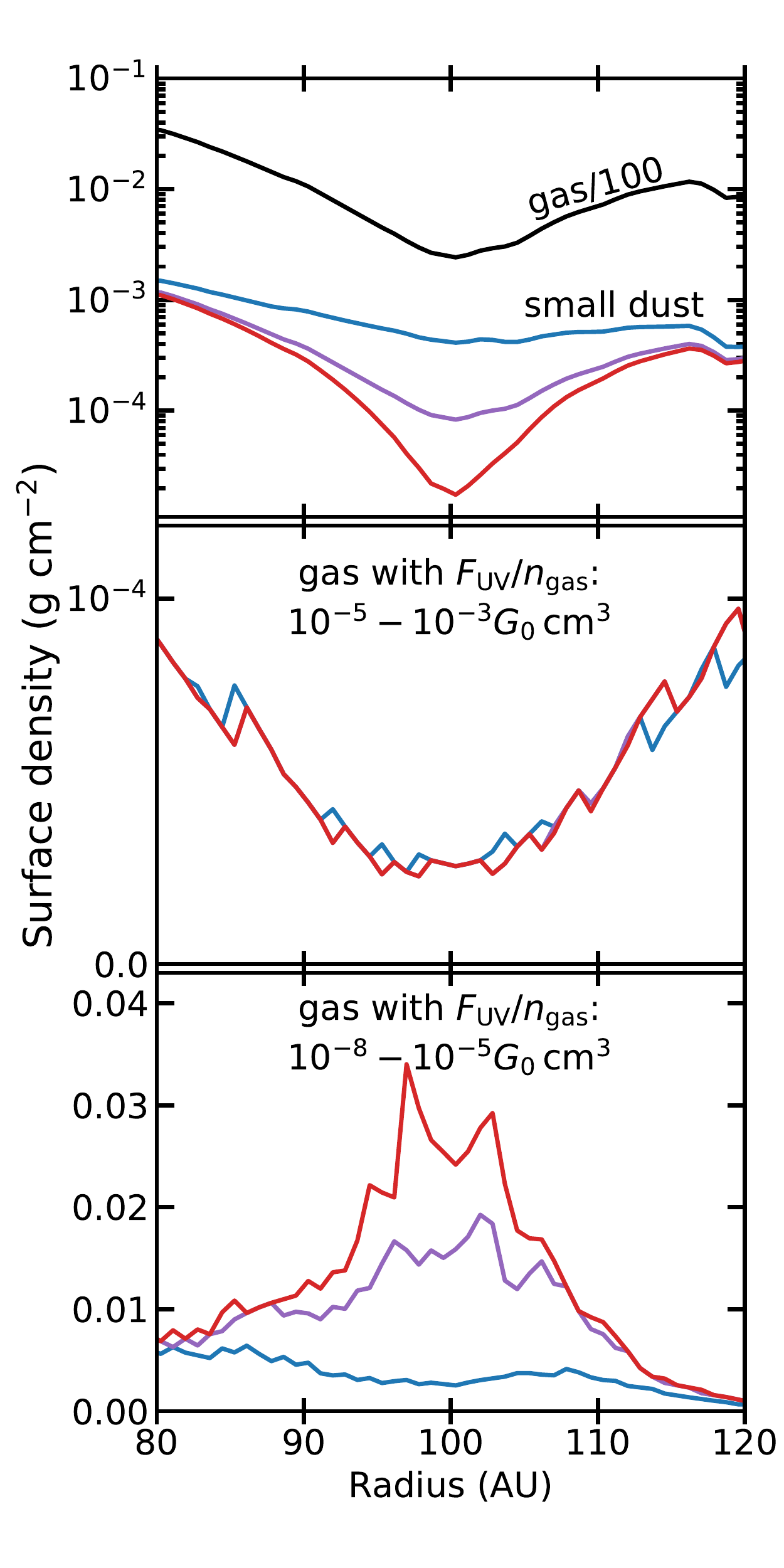}
    \caption{Gas (scaled down by a factor 100) and small dust surface densities (top), and gas surface density that has a $F_\mathrm{UV}/n_\mathrm{gas}$ between $10^{-5}$ and $10^{-3}$ (middle), and  $10^{-8}$ and $10^{-5}$ $G_0\,\mathrm{cm}^{3}$ (bottom) for the 100 AU gap of the AS 209 model of \citet{Alarcon2020}. The small dust surface density is varied in the gap, with a gap that is shallower in the small dust than in the gas (blue), has equal depth in the gas and the small dust (purple) and is deeper in the small dust than in the gap (red). The surface density in the high $F_\mathrm{UV}/n_\mathrm{gas}$ layer follows radial profile of the gas density and does not depend on the amount of small dust. The lower $F_\mathrm{UV}/n_\mathrm{gas}$ layer however is dependent on the levels of small dust depletion and this layer contains more mass, and thus is brighter in \ce{C2H} at lower dust surface densities. }
    \label{fig:AS209_gaps}
\end{figure}

Axisymmetric structures in and above the pebble disks have been observed in many disks in the (sub-)millimeter \citep[e.g.][]{Andrews2018} and scattered light \citep[e.g.][]{Garufi2017}. These features are attributed to changes in the physical conditions that impact the distribution of dust and thus the penetration of UV photon, which directly effects the \ce{C2H} abundance. 

Locations of prime interest in this regard would be planet created gaps. \citet{Alarcon2020} looked at the chemistry in the gaps of AS 209 and found little to no effect of the inclusion of a gap on the \ce{C2H} columns and fluxes for a solar C/O ratio. These models only considered a constant gas-to-small-dust ratio in the gap, while increasing this ratio would enhance the UV penetration and increase the amount of mass available at a given $F_{\mathrm{UV}}/ n_\mathrm{gas}$. 

To check the effects of a possible gap on the distribution of $F_{\mathrm{UV}}/ n_\mathrm{gas}$ in the gap region three models were run, based on the AS 209 model of \citet{Alarcon2020} (See their Table 1. for general model parameters). The model includes a gap, centered around 100 AU. The gap is modeled as a Gaussian with a FWHM of 16 AU and has a factor 12 depletion in the gas and large dust. For the small dust we consider here three different distributions. A model in which the small dust is depressed by the same value as the gas, keeping a constant gas-to-dust ration in the gap. As model in which the dust is only depleted by 20\% w.r.t. a smooth model, having a factor $\sim$ 10 lower gas-to-dust ratio in the gap center, and a model with a small dust depression that is 10 times stronger than than the depression in the gas surface density, leading to a higher gas-to-dust ratio in the gap. The gas and small dust surface densities as well as the mass that is in the strongly irradiated, and mildly irradiated layer in the vicinity of the gap is show in Fig.~\ref{fig:AS209_gaps}.
Interestingly, changing the gas-to-small-dust does not change the amount of mass in the strongly irradiated layer, $F{_\mathrm{UV}}/n{_\mathrm{gas}} = 10^{-5} - 10^{-3}$ $G_0\,\mathrm{cm}^3$. 

\begin{figure*}
    \centering
    \includegraphics[width = \hsize]{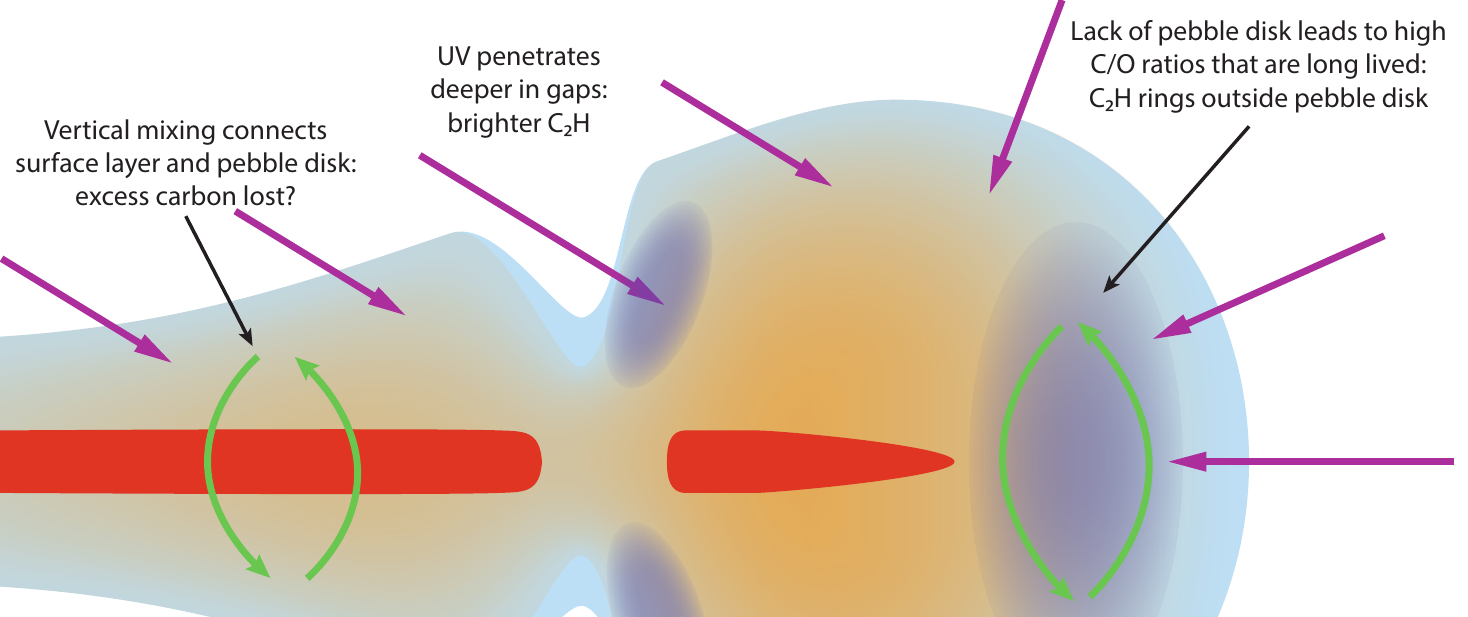}
    \caption{Sketch of the physical processes that set the distribution of \ce{C2H} in the disk. In the inner disk vertical mixing will lock any excess carbon in the mid-plane. We predict that in the disk gap, the increased UV field, combined with the lack of large grains in the mid-plane will} lead to a ring of \ce{C2H} emission. Similarly, the lack of large grains outside of the pebble disk lead to the formation of a large reservoir of long lived carbon rich gas, creating a \ce{C2H} emission ring outside of the pebble disk, such as observed in TW Hya and DM Tau \citep{Bergin2016}.
    \label{fig:overviewsketch}
\end{figure*}

The mass in the mildly irradiated layer, $F{_\mathrm{UV}}/n{_\mathrm{gas}} = 10^{-8} - 10^{-5}$ $G_0\,\mathrm{cm}^3$, does show dependence on the small-dust surface density. Thus gaps with C/O $>$ 1.0 will have an increased column density as long as there is a depletion of the small dust in the gap. We note however that a factor 5 depletion in the small dust only leads to an increase of a factor two in the available mass in these models. This indicates that large variations in the \ce{C2H} column are more likely due to changes in C/O ratios, but that factor few changes in \ce{C2H} column can still be directly due to changes in UV penetration.  

Gaps have an increased UV field in deeper layers of the disk. Aside from the effect that this has on the chemistry directly through changing the distribution $F{_\mathrm{UV}}/n{_\mathrm{gas}}$ it also exposes more and unprocessed dust grains. Photo-desorption and -ablation can then release species from the dust surfaces into the gas. If the grains are water-ice poor, this can increase the C/O ratio, and thus increase the \ce{C2H} abundance. Meridional flows also be efficient in bringing unprocessed material into regions of high UV flux \citep{Morbidelli2014, Teague2019}. This could lead to \ce{C2H} rings around the millimeter dust gaps that have been imaged. Hints of this can be seen in the DM Tau disk \citep{Bergin2016}, although conversely, the rings in TW Hya do not show corresponding \ce{C2H} rings. The \ce{C2H} emission in \citet{Bergner2019} does show some hints of structure, however the resolution of the data is not enough to correlate it with the location of millimeter structure.  Further, high resolution of \ce{C2H} are thus needed to check for a milli-meter gap versus \ce{C2H} ring location correlation in a large sample of disks.

\subsection{Source of Refractory Carbon}
\label{ssc:carbon_carrier}
So far it has been assumed that the source of the excess carbon is on grains in the form of a hydrogenated amorphous carbon. This is however not the only source of refractory carbon in the disk. Up to 20\% of the refractory carbon in the ISM is in the form of PAHs \citep[][]{Tielens2008}. These PAHs could thus provide a significant amount of carbon to the gas, if they are present and can be efficiently destroyed. It is clear from observations that PAHs are not present above the 10 $\mu$m continuum disk photo-sphere \citep{Geers2007}, which means that PAHs are not present in the \ce{C2H} emitting layers. If PAHs act like a large molecule, with corresponding freeze-out behavior, then they should be depleted together with the volatiles and sequestered into the mid-plane. If this is not the case, then the PAH absence needs to be explained by the destruction of PAHs. 

UV photons, especially in T-Tauri disks have difficulties destroying PAHs \citep{Visser2007} so they can be ruled out as a reason for the lack of PAHs in the disk surface. X-rays are, however able to destroy PAHs around T-Tauri stars \citep[e.g.][]{Siebenmorgen2010, Siebenmorgen2012}. In this case, the disk surface layers can easily be enriched in carbon for T-Tauri disks, but for the disks around the X-ray weaker Herbig Ae/Be stars this might not be the case. At the same time, a large reservoir of gas outside the pebble disk near the mid-plane is shielded from X-rays \citep{Rab2018}. This is the region however, that can contribute significantly to the \ce{C2H} emission rings outside of the sub-millimeter continuum. It is thus unlikely that PAH destruction by X-rays can explain all the necessary excess carbon.

As such, to elevate the C/O ratio it is critical that there is enough carbonaceous material in small grains in the outer disk. Collision experiments with carbonaceous grains are scarce, however, there is evidence that carbonaceous grains stick less efficiently than silicate grains at temperatures below 220 K \citep{Kouchi2002}. However, no experiments at cryogenic temperatures have been performed. If the sticking of carbonaceous material is lower than that for silicate materials, would naturally lead to a high fraction of small carbonaceous grains in the surface layers of disk. This could lead to an observable increase of carbonaceous material in the lower density regions of proto-planetary disks and could even help explain the low carbon content of carbonaceous chondrites.

\section{Conclusions}

We have studied the chemistry of \ce{C2H} in the photon-dominated layers of proto-planetary disks. To be able to explain the observed \ce{C2H} we find a specific set of conditions that need to be satisfied. These are also shown schematically in Fig.~\ref{fig:overviewsketch}. 

\begin{enumerate}
    \item The short chemical timescales of \ce{C2H} necessitate a gas-phase equilibrium cycle.
    \item As previously concluded by, for example \citet{Bergin2015} and \citet{Miotello2019}, this cycle is active in disk regions with a C/O ratio $> 1.5$. These high C/O ratios allow the \ce{C2H} emitting layer to extend deeper, from just the top $F_\mathrm{UV}/n_\mathrm{gas} = 10^{-5} - 10^{-3}$ layer at C/O $<$ 1.0, down to the $F_\mathrm{UV}/n_\mathrm{gas} = 10^{-8} - 10^{-3}$ layer. 
    \item To replicate the elevated C/O ratio we propose the photo-ablation refractory carbon carried by carbonaceous grains as the source of the excess carbon in the gas. As significant amounts of oxygen carried by \ce{CO} and \ce{H2O} are no longer present in the disk surface layers, only of 1--10\% of the refractory carbon is necessary to increase the C/O ratio $>$1.5. Based on the existing laboratory data the time-scale for photo-ablation are $< 1$ Myr for the \ce{C2H} emitting area. 
    \item These processes reach a maximum in areas where the small grain population dominates in the mid-plane, which naturally occur at the edges of pebble disks, as well as possibly at the locations of the millimeter gaps, leading to the formation of long-lived rings rich in hydrocarbons. 
\end{enumerate}

Chemical models coupled with gas-grain dynamics can be used to test the efficiency of carbon photo-ablation in increasing the C/O ratios. This vertical motion of the grains and gas needs to be accounted for to assess the fraction of grain that can lose their carbon during the disk life-time and the longevity of the high C/O ratios. These models are out of the scope of this paper, but will be considered in follow-up studies. 

Furthermore, future infrared scattered light and absorption studies will be critical in constraining the abundance as well as the composition of the small grains. Especially absorption studies of edge-on protoplanetary disks should show the absence or presence of the C-H stretch of hydrogenated amorphous carbon and PAHs around 3 $\mu$m.

\acknowledgments
ADB and EAB acknowledge support from NSF Grant\#1907653 and NASA grant XRP 80NSSC20K0259. K. Z. acknowledges the support of NASA through Hubble Fellowship grant HST-HF2-51401.001 awarded by the Space Telescope Science Institute, which is operated by the Association of Universities for Research in Astronomy, Inc., for NASA, under contract NAS5-26555.

\software{RAC2D \citep{Du2014}, SciPy \citep{Virtanen2020},  NumPy \citep{van2011numpy}, Matplotlib \citep{Hunter2007}.}

\bibliographystyle{aa.bst}
\bibliography{Lit_list}

\end{document}